\documentclass[%
reprint,
superscriptaddress,
showpacs,preprintnumbers,
amsmath,amssymb,
aps,
]{revtex4-1}

\usepackage{bmpsize}
\usepackage{color}
\usepackage{graphicx}%
\usepackage{dcolumn}
\usepackage{bm}
\usepackage{physics}

\makeatletter
\def\@hangfrom@section#1#2#3{\@hangfrom{#1#2}#3}
\def\@hangfroms@section#1#2{#1#2}
\makeatother

\begin{document}

\preprint{}

\title{Single Crystal Growth and Transport Properties of van der Waals Materials \textit{AB}Te$_\mathbf{4}$ (\textit{A}/\textit{B} = Ti, Zr, Hf)}

\author{Yuto Hasuo}

\author{Takahiro Urata}
\affiliation{Department of Materials Physics, Nagoya University, Nagoya 464-8603, Japan}

\author{Masaaki Araidai}
\affiliation{Department of Materials Physics, Nagoya University, Nagoya 464-8603, Japan}
\affiliation{Institute of Materials and Systems for Sustainability, Nagoya University, Nagoya 464-8601, Japan}

\author{Yuji Tsuchiya}

\author{Satoshi Awaji}
\affiliation{High Field Laboratory for Superconducting Materials, Institute for Materials Research, Tohoku University, Sendai 980-8577, Japan}

\author{Hiroshi Ikuta}
\affiliation{Department of Materials Physics, Nagoya University, Nagoya 464-8603, Japan}
\affiliation{Research Center for Crystalline Materials Engineering, 
Nagoya University, Nagoya 464-8603, Japan}

\date{\today}

\begin{abstract}
	Monolayers of $AB$Te$_4$ ($A$/$B$ = Ti, Zr, Hf) were theoretically predicted to be two-dimensional topological insulators, but little has been known about the physical properties of these compounds. 
    Here, we report on the single crystal growth, bulk transport properties, and band structure calculations of these compounds.
	The magnetotransport properties indicate that all three compounds are multi-carrier systems.
	The experimental results of ZrTiTe$_4$ and HfTiTe$_4$ can be well fitted by the multi-carrier formula assuming two types of carriers, while three carrier components were necessary for HfZrTe$_4$.
	Interestingly, one of the carrier mobilities of HfZrTe$_4$ exceeded 1000 cm$^2$/V\,s, which was nearly one order in magnitude larger than the carrier mobilities of ZrTiTe$_4$ and HfTiTe$_4$.
	Our band structure calculations showed that all three compounds are semimetals consistent with the magnetotransport properties.
	The band structure around the $\Gamma$-point of HfZrTe$_4$ exhibits features that are distinct from the other two compounds, which is likely the reason of the different carrier properties.
\end{abstract}

\maketitle
\section{Introduction}
Transition metal dichalcogenides MX$_2$ have attracted continuous research interest over the past several decades.
These materials are characterized by their layered structures that are weakly bonded by van der Waals interaction, giving rise to remarkable quasi-two-dimensional properties.
The chemical composition typically involves transition metals from, but not limited to, groups 4, 5, and 6 elements occupying the M sites, while chalcogens such as S, Se, and Te occupy the X sites. 
The diverse combinations of elements, coupled with the multiple polytypes and the strongly anisotropic character, lead to the emergence of a wide range of intriguing physical properties. 
The discovery of charge density waves in 1T-TaS$_2$ in 1974 \cite{Wilson1974,Wilson1975} was one of the most significant events that brought attention to these materials, but numerous other fascinating physical properties have also been found, including superconductivity \cite{Bulaevskii1975}, the Mott insulator state \cite{Sipos2008}, non-saturating giant magnetoresistance effect \cite{Ali2014}, and topologically non-trivial phases \cite{Soluyanov2015,Paul2023}.
In recent years, there is also a rapidly growing research trend exploiting the strong two-dimensional (2D) nature of these materials for potential applications in 2D electronics, and novel quantum phases have been discovered in monolayer samples \cite{Splendiani2010,Wang2012twoD,qian2014quantum}. 

Naturally, the quest for novel physical properties has also been extended to materials composed of more than two elements. 
For instance, the ternary chalcogenides $AB$Te$_4$, where $A$ = Nb, Ta and $B$ = Ir, Rh, were theoretically predicted to be type-II Weyl semimetals \cite{Koepernik2016,Li2017,Liu2017}, which sparked off experimental investigations into these compounds \cite{Haubold2017,Zhou2018,Schoenemann2019,Zhou2019,Shipunov2021,Wang2022}. 
More recently, Macam \textit{et al.} predicted that monolayers of three $AB$Te$_4$ compounds, namely $A$/$B$ = Ti, Zr, or Hf, are 2D topological insulators (TIs) \cite{macam2021tuning}.
Two-dimensional TIs are expected to exhibit intriguing properties not present in three-dimensional (3D) TIs, such as one-dimensional helical edge states, quantized conductance, and the quantum spin Hall effect.
However, while many 3D TIs have been discovered in recent decades \cite{hsieh2008topological,taskin2009quantum,xia2009observation,butch2010strong,kuroda2010experimental,hasan2010colloquium,rachel2018interacting}, the number of known 2D TIs remains limited \cite{bernevig2006quantum,konig2007quantum,roth2009nonlocal,liu2008quantum,du2015robust,Murakami2006,Sabater2013,Drozdov2014,qian2014quantum,tang2017quantum,jia2017direct,wu2018observation}.
Therefore, the prediction of Macam \textit{et al}. is of particular interest, as it may introduce new members to the 2D TI family.

Nevertheless, experimental knowledge about the properties of $AB$Te$_4$ with $A$/$B$ = Ti, Zr, or Hf is very limited as they received only a little attention in the past \cite{cybulski1989structure}.
Even the electronic properties of the bulk state remain unknown.
Furthermore, to the best of our knowledge, there have been no reports of single crystal synthesis for these materials. 
In this work, therefore, we studied the crystal growth of these materials and obtained sizable crystals. 
We then carried out magnetotransport measurements using the single crystals.
Through detailed analysis of the magnetoresistance and Hall resistivity data, we gained insights into the electronic structures that should serve as the basis for further studies on these materials. 
We also performed first-principles calculations and discussed the correlation between the band structure and the experimental results.

\begin{figure}
	\includegraphics[width=\linewidth]{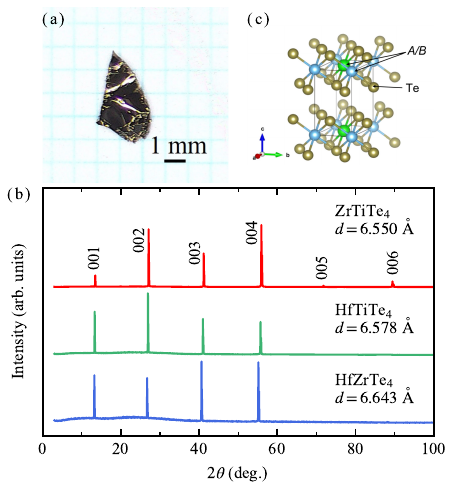}
	\caption{
		\label{fig1}(Color online) (a) Optical image of a HfZrTe$_4$ single crystal.
	(b) Out of plane x-ray diffraction patterns of $AB$Te$_4$.
    (c) Crystal structure of $AB$Te$_4$ ($A$/$B$ = Ti, Zr, Hf) drawn by VESTA \cite{VESTA}.
}
\end{figure}

\section{Methods}
Single crystals of $AB$Te$_4$ ($A$/$B$ = Ti, Zr, Hf) were grown by a chemical vapor transport (CVT) method.
First, polycrystalline precursor powders were prepared by heating stoichiometric amounts of the constituting elements at 900$^\circ$C for 24 hours.
Subsequently, the precursor was sealed into another quartz tube with a transport agent (I$_2$) and was heated in a tube furnace under a temperature gradient.
X-ray diffraction (XRD) and energy dispersive x-ray spectroscopy (EDX) measurements were performed to evaluate the quality of the obtained samples.
Electronic transport measurements with magnetic fields ($H$) up to 9 T were carried out with a Quantum Design Physical Properties Measurement System (PPMS).
In addition, high-field measurements up to 18 T were made at 4.2 K with the superconducting magnet at the High Field Laboratory for Superconducting Materials, Institute for Materials Research (IMR), Tohoku University.
The electrical current was sourced to flow parallel to the (001) plane, and the magnetic field was applied perpendicular to that plane.
The magnetotransport data were measured for both positive and negative magnetic fields, and were symmetrized to compensate for the influence of slight misalignment between the voltage contacts. 
To avoid sample degradation, handling in air was kept to a minimum (within about 1 hour) during the preparation of these measurements.

We also performed density functional theory calculations of bulk $AB$Te$_4$ using the Synopsys QuantumATK version T-2022.03 \cite{QuantumATK}. The optimized norm-conserving Vanderbilt pseudopotentials \cite{schlipf2015optimization} and the Heyd-Scuseria-Ernzerhof hybrid exchange-correlation functional \cite{heyd2006erratum} were used in the calculations. The cutoff energy for space discretization was set to 2721 eV. The Brillouin zone of the unit cell was sampled using a $6\times 10 \times 6$ $\Gamma$-center grid. Structure optimizations were performed by fixing the lattice constants $c$ to the experimental values, and the electronic band structures were calculated for the obtained structural parameters.

\section{Results and Discussion}

\subsection{Crystal growth and characterization}
We first tried to grow single crystals by placing the source materials at the high temperature end of the ampoule, but saw no crystal growth on the low temperature side.
Instead, small crystals were obtained at the high temperature side, indicating that $AB$Te$_4$ grows with an exothermic reaction, similar to HfTe$_2$ \cite{mangelsen2020hfte}.
Hence, the source materials were placed at the colder end of the ampoule to obtain larger crystals. 
The source and growth temperatures were set to 750$^\circ$C and 930$^\circ$C, respectively, and thin plate-like crystals were obtained.
The largest crystal had a size exceeding 5 mm per side.
Figure \ref{fig1}(a) shows an optical microscope image of a HfZrTe$_4$ crystal obtained by this method.

Figure \ref{fig1}(b) shows the out-of-plane XRD patterns of the $AB$Te$_4$ crystals.
Only $00l$ peaks were observed for all three compounds, indicating that the natural facet was the (001) plane.
The crystal structure of ZrTiTe$_4$ is shown in Fig. \ref{fig1}(c) and has a monoclinic structure ($P2/m$), the angle $\beta$ between the $a$ and $c$ axes slightly deviating from 90$^\circ$ \cite{cybulski1989structure}.
Although there are no reports on the structures of HfTiTe$_4$ and HfZrTe$_4$, we expect them to belong to the same crystal system.
Therefore, the interlayer spacing $d$ calculated from the $00l$ peak positions are not strictly the lattice constant $c$.
However, if we calculate $c$ from $d$ using $\beta$ reported for ZrTiTe$_4$ \cite{cybulski1989structure},  the two parameters match within the digits denoted in Fig. \ref{fig1}(b).
Therefore, we regard the values indicated in Fig. \ref{fig1}(b) as the lattice parameters $c$ within the precision of the present experiments.
The composition ratios measured by EDX were Zr : Ti : Te = $1.01\pm 0.04 : 1.03\pm 0.08 : 3.96\pm 0.12$ for ZrTiTe$_4$, Hf : Ti : Te = $1.09\pm 0.03 : 1.04\pm 0.04 : 3.87\pm 0.07$ for HfTiTe$_4$, and Hf : Zr : Te = $1.08\pm 0.07 : 0.99\pm 0.09 : 3.92\pm0.06$ for HfZrTe$_4$.
Although Te was somewhat deficient in all three compounds, the chemical ratios were close to $1:1:4$.

\subsection{Transport properties}

\begin{figure}
\includegraphics[width=\linewidth]{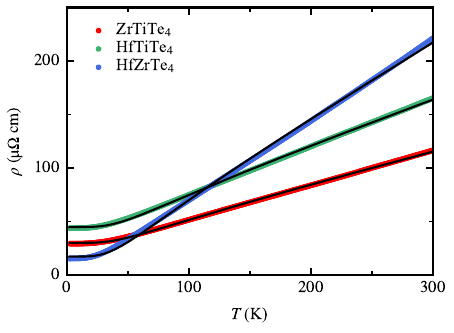}
\caption{\label{resistivity}(Color online) Temperature dependence of resistivity of $AB$Te$_4$. 
The black lines are the results of Bloch-Gr\"{u}neisen fitting.
}
\end{figure}

Figure \ref{resistivity} shows the temperature dependence of resistivity of $AB$Te$_4$.
All three compounds showed a metallic behavior.
The temperature dependence of resistivity is described by the following Bloch-Gr\"{u}neisen formula if electron-phonon scattering processes are dominant \cite{ziman1972principles}: 
\begin{equation}
  \rho\ \qty(T)=\rho_0 +A \qty(\frac{T}{\Theta_\mathrm{D}})^5 \int_{0}^{\Theta_\mathrm{D}/T} \frac{x^5}{\qty(e^x-1)(1-e^{-x})}\dd{x},
\end{equation}
where $\rho_0$ is the residual resistivity and $\Theta_\mathrm{D}$ the Debye temperature.
We fitted this formula to the data, and the results are indicated by black curves in the figure.
Good agreement with the experiments was found for all compounds, indicating that the resistivity is governed by electron-phonon scattering below 300 K.
The Debye temperatures derived from the fittings were $243.8 \pm 1.6$ K for ZrTiTe$_4$, $243.7 \pm 1.5$ K for HfTiTe$_4$, and $221.7 \pm 2.2$ K for HfZrTe$_4$.

\begin{figure*}
	\includegraphics[width=\linewidth]{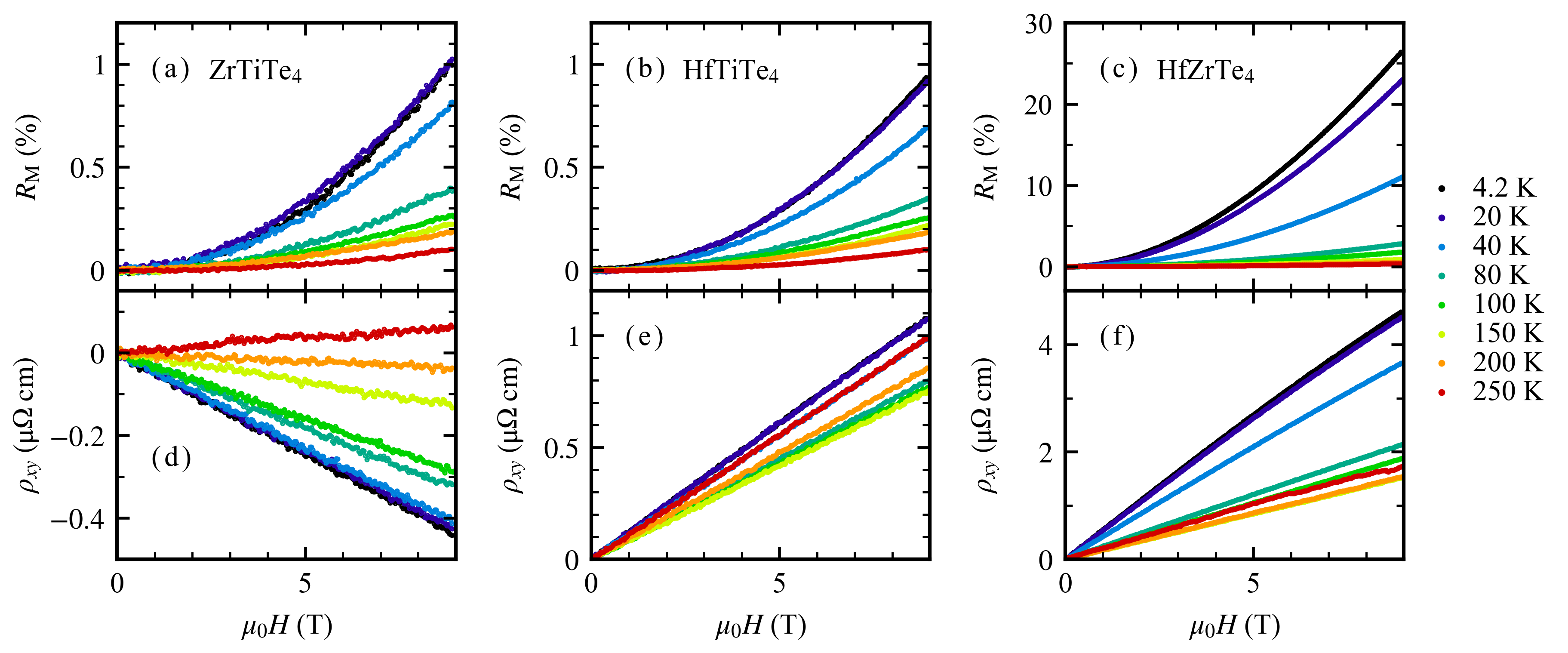}
	\caption{\label{MR}(Color online) Magnetic field dependencies of magnetoresistance ratio ($R_\mathrm{M}$) and Hall resistivity measured up to 9 T at various temperatures for (a), (d) ZrTiTe$_4$, (b), (e) HfTiTe$_4$, and (c), (f) HfZrTe$_4$. 
}
\end{figure*}

Figures \ref{MR}(a)-\ref{MR}(c) show the magnetic field dependence of magnetoresistance ratio measured at various temperatures up to 9 T.
The magnetoresistance ratio $R_\mathrm{M}$ was defined as $R_\mathrm{M} = 100\% \times [\rho_{xx}(\mu_0 H) - \rho_{xx}(0)]/\rho_{xx}(0)$.
A positive magnetoresistance that increased monotonically with the magnetic field was observed for all compounds.
$R_\mathrm{M}$ increased with decreasing temperature, reflecting the increase in mobility due to the suppression of electron-phonon scattering.
Figures \ref{MR}(d)-\ref{MR}(f) show the magnetic field dependence of Hall resistivity measured at various temperatures.
For all three compounds, the field dependence was fairly linear up to 9 T for most of the measured temperatures, although a small deviation from linearity was notable for HfZrTe$_4$ at low temperatures.
The field gradient of the Hall resistivity was negative for ZrTiTe$_4$ except at 250 K, whereas it was positive at all temperatures for HfTiTe$_4$ and HfZrTe$_4$.
These results indicate that the dominant carriers are electrons in ZrTiTe$_4$, while they are holes in the latter two.

\begin{figure}
	\includegraphics[width=\linewidth]{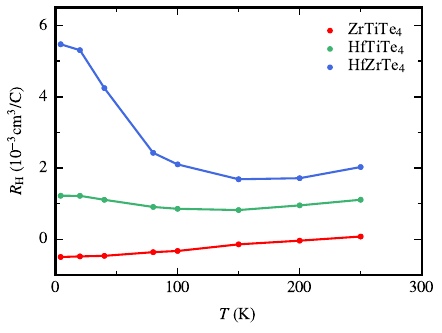}
	\caption{\label{hallcoef}(Color online) Temperature dependence of Hall coefficient ($R_\mathrm{H}$).
	$R_\mathrm{H}$ was determined from the slope of Hall resistivity in the range of $0\leq \mu_0 H\leq 3$ T.
}
\end{figure}

Figure \ref{hallcoef} shows the temperature dependence of Hall coefficient ($R_\mathrm{H}$) determined from the slope of Hall resistivity in the range of $0\leq \mu_0 H\leq 3$ T.
$R_\mathrm{H}$ depended on temperature for all three compounds, and the absolute value increased in the order of ZrTiTe$_4$, HfTiTe$_4$, to HfZrTe$_4$. 
While the temperature dependence of $R_\mathrm{H}$ was monotonic for ZrTiTe$_4$, it was evidently non-monotonic for the other two compounds.
The fact that $R_\mathrm{H}$ depended on temperature can be attributed to multiband effects, consistent with the multi-carrier fitting to the magnetotransport data discussed below. 

\begin{figure}
	\includegraphics[width=\linewidth]{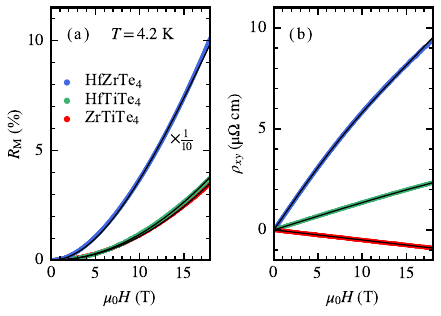}
	\caption{\label{MR_18T}(Color online) Magnetic field dependencies of (a) magnetoresistance ratio ($R_\mathrm{M}$) and (b) Hall resistivity measured up to 18 T at 4.2 K using the superconducting magnet at IMR, Tohoku University. 
	For better visualization, the $R_\mathrm{M}$ data of HfZrTe$_4$ were divided by a factor of 10 in this plot.
	The black lines indicate the results of multi-carrier fittings. }
\end{figure}

\begin{table*}
  \begin{center}
	\caption{\label{params}Carrier densities and mobilities of $AB$Te$_4$ determined from the multi-carrier fitting to the magnetotransport data of Fig. \ref{MR_18T}.
	$n$ $(p)$ is the carrier density and $\mu$ $(\nu)$ is the mobility of electrons (holes).}
		\begin{ruledtabular}
    \begin{tabular}{ccccccc}
          & $n\, (\mathrm{cm}^{-3})$ & $p_1\, (\mathrm{cm}^{-3})$ & $p_2\, (\mathrm{cm}^{-3})$ & $\mu\, (\mathrm{cm^{2}/V\, s})$ & $\nu_1\, (\mathrm{cm^{2}/V\, s})$ & $\nu_2\, (\mathrm{cm^{2}/V\, s})$ \\
        \hline 
				ZrTiTe$_4$           & $6.16\times 10^{20}$ & $6.73\times 10^{20}$  & - & 112                  & 96.4          &  -     \\
				HfTiTe$_4$           & $1.16\times 10^{21}$ & $4.56\times 10^{20}$  & - & 68.1                 & 149           &  -    \\			 
        HfZrTe$_4$          & $3.42\times10^{20}$ & $5.31\times10^{20}$   & $3.49\times10^{20}$   & 414                         & 202                          & 1101   \\                        
    \end{tabular}
	\end{ruledtabular}
  \end{center}
\end{table*}

To gain insight into the carrier properties, we fitted the multi-carrier model described below to the magnetotransport data taken at 4.2 K.
However, there remained some ambiguity in the results, in particular the fit to the HfZrTe$_4$ data did not converge to a unique parameter set.
To improve the quality of the fit, therefore, we measured the magnetotransport properties up to higher fields (up to 18 T) using the superconducting magnet at IMR, Tohoku University, the results of which are plotted in Fig. \ref{MR_18T}.
The low-field parts ($\mu_0H \leq 9$ T) of these data agreed well with the data shown in Fig. \ref{MR}.
From Fig. \ref{MR_18T}(a), we see that the magnetoresistance ratio is quite different between HfZrTe$_4$ and the other two compounds.
While $R_\mathrm{M}$ was around 3-4\% at 18 T for ZrTiTe$_4$ and HfTiTe$_4$, it reached 100\% for HfZrTe$_4$, suggesting that there is a qualitative difference in the electronic structures between the latter and the other two compounds.
The Hall resistivity of ZrTiTe$_4$ showed an almost linear field dependence up to 18 T as shown in Fig. \ref{MR_18T}(b), maintaining the trend we noticed above for the data up to 9 T.
For HfTiTe$_4$, on the other hand, a slight non-linearity became apparent by extending the field range to 18 T, and the non-linear field behavior of HfZrTe$_4$, which was already noticeable in magnetic fields up to 9 T, was even more pronounced.

Under the relaxation time approximation, the conductivity tensor components $\sigma_{xx}$ and $\sigma_{xy}$ are given by \cite{kim1993multicarrier}
\begin{align}
\sigma_{xx}&= \sum_{i} \frac{en_i \mu_i}{1+(\mu_i B)^2}\label{s_xx},\\
\sigma_{xy}&= \sum_{i} \frac{s_ien_i \mu_i^2B}{1+(\mu_i B)^2}\label{s_xy},
\end{align} 
where subscript $i$ indicates the $i$-th carrier component, $e$ is the elementary charge, $n_i$ carrier density, $\mu_i$ mobility, and $s_i$ takes $-1$ for electron-type carriers and $+1$ for hole-type carriers.
The measured quantities $\rho_{xx}$ and $\rho_{xy}$ were converted to conductivity components $\sigma_{xx}$ and $\sigma_{xy}$ by the relations $\sigma_{xx} = \rho_{xx}/(\rho_{xx}^2 + \rho_{xy}^2)$ and $\sigma_{xy} = \rho_{xy}/(\rho_{xx}^2 + \rho_{xy}^2)$.
Equations (\ref{s_xx}) and (\ref{s_xy}) were simultaneously fitted to the experimental $\sigma_{xx}$ and $\sigma_{xy}$ data with the carrier density ($n_i$) and the mobility ($\mu_i$) as free parameters.

We first assumed a single carrier system, but the fit did not give a good approximation of the data.
It turned out that a two-carrier model with one electron- and one hole-type carriers described well the data of ZrTiTe$_4$ and HfTiTe$_4$.
For HfZrTe$_4$, however, the fitted result was still not satisfactory.
Therefore, we fitted the data with three carrier components and found that assuming one electron- and two hole-type carriers yielded a good agreement with the data.
The black lines in Fig. \ref{MR_18T} show the fitted curves.
The reason of the difference in the numbers of the carrier components will be discussed in the following subsection.

Table \ref{params} summarizes the results of the multi-carrier fitting.
All three compounds have relatively large carrier densities $10^{20}$-$10^{21}$ $\mathrm{cm}^{-3}$, which is consistent with the metallic temperature dependence of resistivity (Fig. \ref{resistivity}).
It is worth mentioning that the carrier mobilities of HfZrTe$_4$ are larger than the other two compounds. 
In particular, it exceeds 1000 cm$^2$/V\,s for one of the hole-type carriers.

From Table \ref{params}, we see that ZrTiTe$_4$ is an almost perfectly compensated semimetal having similar electron and hole densities.
In addition, the mobilities are also close to each other.
The Hall coefficient in the low-field limit of a semimetal that has one electron- and one hole-type carriers can be written as \cite{ziman1972principles}
\begin{align}
	R_\mathrm{H} = \frac{n_h\mu_h^2 - n_e\mu_e^2}{e(n_h\mu_h + n_e\mu_e)^2}\label{eq_hall},
\end{align}
where $n_e$ ($n_h$) and $\mu_e$ ($\mu_h$) are electron (hole) density and mobility, respectively.
It is easy to see from this equation that the Hall coefficient of ZrTiTe$_4$ should have a small magnitude. 
Moreover, the electrons have a slightly larger mobility than the holes in ZrTiTe$_4$, which explains the negative sign of the Hall coefficient at low temperatures. 
In fact, $R_\mathrm{H}$ calculated using the carrier parameters of ZrTiTe$_4$ summarized in Table \ref{params} is $-5.13\times 10^{-4}$ cm$^3$/C, which is close to the low-field Hall coefficient at 4.2 K shown in Fig. \ref{hallcoef}.
On the other hand, both the densities and mobilities of the two carrier components of HfTiTe$_4$ are certainly different to each other, and the mobility of the holes is larger than that of the electrons.
Thus, the Hall coefficient is positive and the absolute value is larger than ZrTiTe$_4$.
Quantitatively, calculating the value of $R_\mathrm{H}$ from the carrier parameters shown in Table \ref{params} gives $1.37\times 10^{-3}$ cm$^3$/C for HfTiTe$_4$, again close to the experimental result of Fig. \ref{hallcoef}.
It is also noteworthy that we observed a slight non-linearity in the field dependence of the Hall resisitivity of HfTiTe$_4$ as mentioned above, which is consistent with the lack of complete compensation between the two types of carriers. 
Finally, the three carrier components of HfZrTe$_4$ have very different characteristics from each other, especially the mobilities are largely different.
Therefore, the contribution from each carrier component would be compensated to a lesser extent, which explains why the magnitude of the Hall coefficient is the largest among the three compounds.

\subsection{Band structure calculation}

\begin{table*}
  \begin{center}
		\caption{\label{lattice}Optimized lattice parameters of $AB$Te$_4$.}
		\begin{ruledtabular}
    \begin{tabular}{ccccccc}
          & $a$ (\AA) & $b$ (\AA)& $c$ (\AA) (exp.) & $\alpha$ (deg.) & $\beta$ (deg.) & $\gamma$ (deg.) \\
        \hline 
				ZrTiTe$_4$           & 6.7963 & 3.9182  & 6.550 & 90.000        & 90.0616          &  90.000     \\
				HfTiTe$_4$           & 6.6892 & 3.8632  & 6.578 & 90.000        & 90.0601          &  90.000     \\
        HfZrTe$_4$           & 6.8661 & 3.9607  & 6.643 & 90.000        & 90.0613          &  90.000     \\              
    \end{tabular}
	\end{ruledtabular}
  \end{center}
\end{table*}

To better understand the experimental results, we also carried out electronic band structure calculations.
Table \ref{lattice} shows the lattice parameters obtained from structural optimizations with the lattice constant $c$ fixed to the experimental value.
As shown in the table, the angles $\beta$ were slightly off from 90$^\circ$ for all three compounds, consistent with the expected monoclinic structure.

\begin{figure*}
	\includegraphics[width = \linewidth]{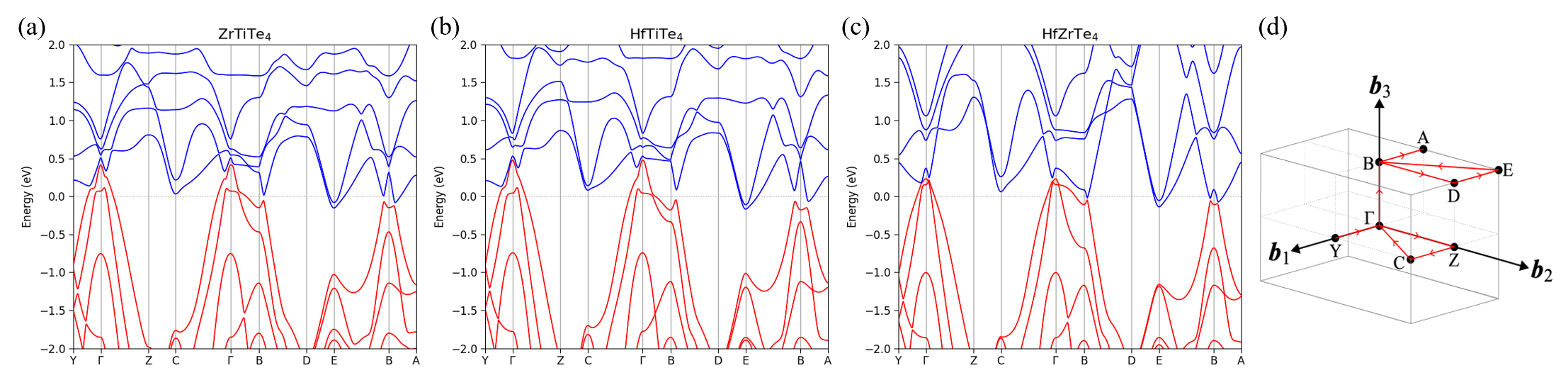}
	\caption{\label{band}(Color online) Electronic band structure of (a) ZrTiTe$_4$, (b) HfTiTe$_4$, and (c) HfZrTe$_4$.
		(d) Brillouin zone of $AB$Te$_4$. 
}
\end{figure*}

Figures \ref{band}(a)-\ref{band}(c) show the electronic band structures of $AB$Te$_4$ calculated using the lattice parameters listed in Table \ref{lattice}, while Fig. \ref{band}(d) depicts the Brillouin zone.
All three compounds have a semimetallic band structure with two hole pockets at the $\Gamma$-point, two electron pockets at the E-point and one electron pocket between the A- and B-points.
The results of our band structure calculations assuming bulk samples showed some similarity to those of monolayer systems \cite{macam2021tuning}.
The band structure of a monolayer system also showed a hole pocket at the $\Gamma$-point and an electron pocket at the S-point (which corresponds to the C-point in Fig. \ref{band}(d)), although the details around the $\Gamma$-point were somewhat different.

While more than two bands cross the Fermi level, the experimental results of ZrTiTe$_4$ and HfTiTe$_4$ were well reproduced by a two-carrier model.
This suggests that the carriers of the same type (electron or hole) have a similar mobility.
As can be seen from Eqs. (\ref{s_xx}) and (\ref{s_xy}), the contribution from carriers of the same type cannot be separated if their mobility are the same.
If several components of the same carrier type have the same mobility, they will behave as if there were only one component with a carrier density equal to the sum of each component.
On the other hand, three components were necessary to fit to the data of HfZrTe$_4$, \textit{i.e.}, one electron-type and two hole-types.
This indicates that the two hole-type components have different mobilities.
Indeed, the mobilities of the two hole-type components of HfZrTe$_4$ obtained by the multi-carrier fitting differ by a factor of 5 as shown in Table \ref{params} ($\nu_1$ and $\nu_2$).
The characteristic difference of the hole-type carriers between HfZrTe$_4$ and the other two compounds can be attributed to the band structure around the $\Gamma$-point.
It can be seen from Figs. \ref{band}(a)-\ref{band}(c) that the band structure around the $\Gamma$ point of ZrTiTe$_4$ and HfTiTe$_4$ is very similar near the Fermi level, but significantly different for HfZrTe$_4$.

\section{Conclusion}
In conclusion, we have grown single crystals of $AB$Te$_4$ ($A$/$B$ = Ti, Zr, Hf) by a CVT method and gained insight into the electronic structure through transport measurements and band structure calculations.
We carried out multi-carrier fittings to the magnetoresistance and Hall resistivity data.
For ZrTiTe$_4$ and HfTiTe$_4$, the two-carrier model assuming one electron- and one hole-type carriers well reproduced the experimental results.
On the other hand, three components, \textit{i.e.}, one electron-type and two hole-type carriers, were necessary to fit the data of HfZrTe$_4$, one of the hole-type carriers having a high mobility exceeding 1000 cm$^2$/V\,s.
Our band structure calculations showed that the three $AB$Te$_4$ compounds all have a semimetallic band structure.
However, the shape of the hole bands around the $\Gamma$-point of HfZrTe$_4$ differs from those of the other two, which is probably the reason of its distinct carrier properties.

\section*{Acknowledgments}
The authors thank Y. Fujisawa, Y. Obata, and Y. Okada for fruitful discussion.
This work was supported by JSPS Grant-in-Aid for JSPS Fellows (Grant No. JP23KJ1124).
The computation was carried out in part using the computer resource offered under the category of
General Projects by Research Institute for Information Technology, Kyushu University.

\bibliography{71041.bib}

\end{document}